\documentclass[aps,pra,reprint, amsmath, amssymb,superscriptaddress,nofootinbib]{revtex4-1}

\usepackage{bm}
\usepackage[retainorgcmds]{IEEEtrantools}
\usepackage{graphicx}
\usepackage{mathrsfs}
\usepackage{amsmath}
\usepackage{amssymb}
\usepackage{color}
\usepackage{amsfonts}
\usepackage{times,txfonts}
\usepackage{nicefrac}
\usepackage[colorlinks=true,linkcolor=blue,urlcolor=blue,citecolor=blue,pdfusetitle]{hyperref}
\usepackage{amsmath}

\begin{document}

\title{Lower bound for entropy production rate in stochastic systems far from equilibrium}
\date{\today}
\author{Domingos S. P. Salazar}
\affiliation{Unidade de Educa\c c\~ao a Dist\^ancia e Tecnologia,
Universidade Federal Rural de Pernambuco,
52171-900 Recife, Pernambuco, Brazil}

\begin{abstract}
We show that the Schnakenberg's entropy production rate in a master equation is lower bounded by a function of the weight of the Markov graph, here defined as the sum of the absolute values of probability currents over the edges. The result is valid for time-dependent nonequilibrium entropy production rates.  Moreover, in a general framework, we prove a theorem showing that the Kullback-Leibler divergence between distributions $P(s)$ and $P'(s):=P(m(s))$, where $m$ is an involution, $m(m(s))=s$, is lower bounded by a function of the total variation of $P$ and $P'$, for any $m$. The bound is tight and it improves on Pinsker's inequality for this setup. This result illustrates a connection between nonequilibrium thermodynamics and graph theory with interesting applications.
\end{abstract}
\maketitle{}


{\bf \emph{Introduction -}} 
Nonequilibrium physics has few general results and most of them can be traced back to the second law of thermodynamics. Among equivalent statements, the second law demands that the entropy production rate of a system is non-negative \cite{RevModPhys.93.035008,Seifert2012,Campisi2011,Bustamante2005,Esposito2009},
\begin{equation}
\label{2ndlaw}
    \frac{d}{dt}\langle \Sigma \rangle = \langle \pi \rangle \geq 0,
\end{equation}
which means that the rate of entropy variation of the system plus environment is non-negative. The statement (\ref{2ndlaw}) is equivalent to all classic statements from Clausius \cite{Clausius1854,Clausius1865}, Carnot \cite{Carnot1824}, and others \cite{RevModPhys.93.035008}, but we used a notation in (\ref{2ndlaw}) suggesting that the quantity $\langle \Sigma \rangle$ is an average. As a matter of fact, with the advent of stochastic thermodynamics \cite{Jarzynskia2008,Jarzynski1997,Jarzynski2000,Crooks1998,Gallavotti1995,Evans1993,Hanggi2015,Saito2008,PhysRevX.11.031064,Gingrich2016}, quantities such as heat, work and even entropy production might be defined at trajectory level. In order to account for fluctuations of thermodynamic quantities in small systems, the second law is now usually rephrased as an ensemble average of the entropy production.

However, for each system, one must identify what plays the role of the entropy production rate in terms of the system's dynamics. 
In several situations, either the entropy production $\langle \Sigma \rangle$ or the entropy production rate $\langle \pi \rangle$ might take the form
\begin{equation}
\label{prodrate}
D(P|P'):=\sum_{s}P(s)\ln\big(\frac{P(s)}{P(s')}\big) \geq 0,
\end{equation}
where $s \in S$ is some event (for instance, a transition), $s'$ is the time reversed event and $P(s)$ is a probability function. This expression in the context of Markov processes \cite{Schnakenberg1976,Tome2012}, heat exchange problems \cite{Hasegawa2019,Timpanaro2019B,Evans2002,Merhav2010,Seifert2005,Jarzynski2004a,Campisi2015} and stochastic thermodynamics \cite{Seifert2012} in general for a suitable choice of $S$ (see applications).

Formally, expression (\ref{prodrate}) is the Kullback-Leibler (KL) divergence of $P(s)$ and $P(s')$. From definition, the expression (\ref{prodrate}) satisfies (\ref{2ndlaw}). Moreover, it vanishes if and only if $P(s)=P(s')$ for all $s$,  which is a form of detailed balance condition. It makes equilibrium a situation equivalent to microscopic reversibility. Actually, expression (\ref{prodrate}) appears as the entropy production rate in the analysis of Lyapunov stability of Markov processes to demonstrate that the system tends to equilibrium in the long time. A nonequilibrium steady state (NESS) can be maintained by coupling the system to multiple reservoirs, resulting in a constant $\langle \pi \rangle>0$.
Thus, $D(P|P')$ quantifies a disagreement the system has with the detailed balance condition (equilibrium). Of course, there are other ways to quantify this disagreement.

Schnakenberg \cite{Schnakenberg1976} proposed a network representation of the master equation, importing results from graph theory \cite{Witthaut2022,GraphTheoryReview,Bollobas1998} to nonequilibrium thermodynamics. 
We explore this idea by considering the network representation of the vertices $s$ for a given matching $\{(s,m(s))|s\in S\}$, where $m(m(s))=1$. We define the weight of this matching as the following sum
\begin{equation}
\label{weight}
w:=\frac{1}{2}\sum_{s}|P(s)-P(s')|,
\end{equation}
which is the total variation distance of $P(s)$ and $P(s')$. As in the case of the entropy production rate (\ref{prodrate}), this weight $w$ is non-negative and $w=0$ if and only if $P(s)-P(s')=0$ for all $s$, which is the detailed balance condition. From definition (\ref{weight}), we also have $w\geq 0$ finite, while $\langle \pi \rangle$ might diverge. Intuitively, the weight of an edge $|P(s)-P(s')|$ quantifies the disagreement with the detailed balance condition on that edge. Summing over all edges results in the amount of disagreement over the whole graph.

A natural question is if both measures of disagreement, $D(P,P')$ and $w$, are somehow connected. For now, we only checked that they agree in equilibrium, $D(P,P')=0$ , $w=0$. In this letter, we show that
\begin{equation}
\label{mainresult}
D(P,P')\geq 2w\tanh^{-1}{w},
\end{equation}
and the bound is tight, valid arbitrarily far from equilibrium, with saturation observed for a particular two-level system. 

The letter is organized as follows. We introduce the formalism and prove a general theorem about the relation between the Kullback-Leibler (KL) divergence \cite{KullbackS.andLeibler1951} and the total variation distance (TV) \cite{Devroye2001} when distributions are related by an involution. We apply this result to the heat exchange problem and comment on the exchange fluctuation theorem. Then, we apply this result to the master equation (\ref{Markov}) and get a bound (\ref{mainresult}) in terms of the probability currents. We also illustrate the result with simulations and compare it with other known bounds for KL and TV from statistics (Pinsker's and Bretagnolle-Hubber's). We apply the theorem to a general setup in stochastic thermodynamics and obtain a lower bound for the stochastic work. Finally, we apply the theorem to a time rotation problem, where a demon flips a coin and decide if time goes forward or backwards. We also discuss some intuition behind the theorem.

{\bf \emph{Formalism -}} First, we define some concepts. Let $P$ and $Q$ be probabilities in $S$, $\sum_s P(s)=\sum_s Q(s) =1$ with $0\leq P(s),Q(s)\leq 1$. The Kullback-Leibler (KL) divergence is given by $D(P|Q):=\sum_s P(s) \ln (P(s)/Q(s))$ (defined when $Q(s)=0 \rightarrow P(s)=0$) and total variation distance, $\Delta(P,Q):=(1/2)\sum_s|P(s)-Q(s)|$. We prove the following result:

{\bf \emph{Theorem -}} {\emph{Let $S\neq \emptyset$ be a countable set and let $P:S\rightarrow [0,1]$ be any probability function. Let $m:S\rightarrow S$ be any involution, $m(m(s))=s$. Define $P':=S\rightarrow [0,1]$ as $P'(s):=P(m(s))$. Then, the bound
\begin{equation}
\label{theorem}
    D(P|P')\geq 2 \Delta(P,P') \tanh^{-1}(\Delta(P,P'))
\end{equation}
is tight.
}}

{\emph{Proof -}} Let $P$ and $m$ be defined in $S$ as in the theorem. If there is an $s \in S$ such that $P(s)=0$ and $P'(s)=P(m(s))\neq 0$, then $D(P|P')$ diverges while $\tanh^{-1}(\Delta(P,P'))$ might be finite. Therefore, we focus on the case with the condition $P(s)=0\rightarrow P(m(s))=0$ for all $s$ (absolute continuity). 

Define a function $\pi:S\rightarrow \mathbb{R}$  as
\begin{equation}
\label{defpi}
    \pi(s):=\ln(P(s)/P(s')),
\end{equation}
for $P(s)>0$, (which makes $P(s')>0$), where $s':=m(s)$. In case $P(s)=0$ (which makes $P(s')=0$), define $\pi(s):=0$. It is immediate that $\pi(s')=\ln(P(m(s))/P(m(s'))=\ln(P(s')/P(s))=-\pi(s)$. Note that the average of $\pi(s)$ is given by
\begin{equation}
\label{checkaverage}
    \langle \pi \rangle := \sum_{s\in S}\pi(s)P(s)=\sum_{s\in S}P(s)\ln\big(\frac{P(s)}{P(s')}\big)=D(P|P').
\end{equation}
Moreover, note that the variable $\pi(s)$ satisfies a form of detailed fluctuation theorem (DFT) from definition (\ref{defpi}), $P(s)/P(s')=\exp(\pi(s))$. Now we define another probability function over the pairs $E=\{e=\{s,m(s)\}|s\in S\}$, $p(e):E \rightarrow [0,1]$, given by
\begin{equation}
    p(e):=P(s)+P(s'),
\end{equation}
if $s\neq s'$ and $p(e):=P(s)$ if $s=s'$.
Because $m$ is an involution, each $s \in S$ belongs to a single pair $e\in E$. We check the normalization of $p(e)$,
\begin{equation}
    \sum_{e\in E} p(e) = \frac{1}{2}\sum_{s,s' \in S} P(s)+P(s')=1,
\end{equation}
and define $\Pi(e):=p(e)^{-1}(P(s)-P(s'))\ln(P(s)/P(s'))$, for the pair $e=\{s,s'\}$, whenever $P(s)>0$. When $P(s)=0$, let $\Pi(e):=0$. Then, we get from (\ref{checkaverage}):
\begin{eqnarray}
\langle \pi \rangle = \frac{1}{2}\sum_{s,s' \in S} (P(s)-P(s'))\ln(\frac{P(s)}{P(s')})=\sum_{e\in E} \Pi(e)p(e).
\end{eqnarray}
We also compute the total variation of probabilities ($P,P'$):
\begin{eqnarray}
\label{TV2}
    \Delta(P,P')=
    \frac{1}{2}\sum_{s \in S}|P(s)-P(s')|= \sum_{e\in E} w(e)p(e),
\end{eqnarray}
where $w(e):=|P(s)-P(s')|/p(e)$, for $e=\{s,s'\}$, when $P(s)>0$ and $w(e):0$, when $P(s)=0$. Finally, for any given pair $e=\{s,s'\}$ with $P(s)>0$, we define $p=\max (P(s),P(s'))/p(e)$ for some $0.5 \leq p < 1$. Then, $w(e)=p-(1-p)$. Also,
\begin{equation}
\label{2lvl}
    \Pi(e)=[p-(1-p)]\ln \big(\frac{p}{1-p}\big)=w(e)\ln \big(\frac{p}{1-p}\big),
\end{equation}
Finally, define the logistic function $\sigma(a)=\exp(a/2)/[\exp(a/2))+\exp(-a/2)]$ and find $a$ such that $p=\sigma(a)$. Then, using $\ln [p/(1-p)]=a$ and $w(e)=p-(1-p)=\tanh(a/2)\rightarrow a = 2\tanh^{-1}(w(e))$ in (\ref{2lvl}), we get
\begin{equation}
\label{2lvlb}
    \Pi(e)= a w(e) = 2w(e)\tanh^{-1}(w(e)):=B(w(e)),
\end{equation}
where $B(w)=2w\tanh^{-1}(w)$. The degenerate case, $P(s)=0 \rightarrow \Pi(e)=0, w(e)=0$ also satisfies (\ref{2lvlb}).
Note that, for $0<w<1$, we have $B(w)>0$, $B'(w)>0$ and $B''(w)>0$. Therefore, from Jensen's inequality,
\begin{equation}
\label{2lvlc}
    \langle \pi \rangle =\sum_{e \in E} \Pi(e) p(e) = \sum_{e \in E} B(w(e))p(e) \geq B\big(\sum_{e \in E} w(e)p(e)\big),
\end{equation}
and, after using $\langle \pi \rangle = D(P,P')$ from (\ref{checkaverage}) and $\sum_e w(e)p(e) = \Delta(P,P')$ from (\ref{TV2}), it results in
\begin{equation}
\label{boundformalism}
    D(P,P')\geq 2\Delta(P,P')\tanh^{-1}(\Delta(P,P')),
\end{equation}
which proves (\ref{theorem}). 

Saturation of the bound follows from (\ref{2lvlb}). If $S$ has a single element, then the bound is trivially saturated $D(P|P')=B(\Delta(P,P'))=0$, as $P=P'$. If S has more than one element, consider two distinct elements from $\{s_1,s_2\}\subseteq S$, and a $\tilde{P}$ such that $\tilde{P}(s_1)+\tilde{P}(s_2)=1$, with involution $\tilde{m}(s_1)=s_2, \tilde{m}(s_2)=s_1$, and $\tilde{m}(s)=s$ otherwise. It makes $p(e)=1$ for $e=\{s_1,s_2\}$ and $\Pi(e)=D(\tilde{P}|\tilde{P}')$, $\Delta(\tilde{P},\tilde{P}')=w(e)$, which results in the saturation, $D(\tilde{P}|\tilde{P}')=B(\Delta(\tilde{P},\tilde{P}'))$. For completeness, for a given total variation $\Delta=\Delta(P,P')$, if there is a lower bound $B^*(\Delta)$ for $D(P,P')$ such that $D(P,P')\geq B^*(\Delta) > B(\Delta)$, then the two level system presented above violates it. Therefore, $B(\Delta)$ is the tightest bound for this setup. That completes the proof.

{\bf \emph{Remarks-}}
Interestingly, time was not mentioned in the theorem and the fluctuation theorem was not assumed. This proof shows the important role played by the involution $m$ in nonequilibrium physics: When you equip a probability space with an involution, you could think of the connected events \{$s,s'$\} as forward and backward events in the arrow of time. Because they might have different probabilities, $P(s)>P(s')$, we pick $s$ as the event forward in time. In this context, the fluctuation theorem is also a construct, because $\pi(s)$ behaves like a thermodynamic force \cite{Gingrich2016}, switching signs after an involution.

Additionally, when compared to off the shelf bounds between KL and TV, we have 
$D(P|P')\geq B(\Delta(P,P'))\geq 2\Delta(P,P)^2$, which shows the bound improves on
Pinsker's inequality \cite{PinskerInequality} for this setup. It is also true for the Bretagnolle-Hubber' \cite{Bretagnolle1979} bound, $D(P|P')\geq B(\Delta(P,P'))\geq -\ln(1-\Delta(P|P')^2)$.

Also note that if $(S_1,P,m)$ satisfies the theorem and $g:S_1 \rightarrow S_2$ is a bijection, then $(S_2, P\circ g^{-1},g\circ m \circ g^{-1})$ satisfies the theorem. For that reason, different systems might be mapped into each other and the bound is universal. For completeness, although the theorem was proved for discrete sets $S$, it is possible to expand it to $S=\mathbb{R}^n$ with a differentiable involution $m$ (see Appendix for the case $S=\mathbb{R}$).

{\bf \emph{Application I: heat exchange problem-}}
In a general heat exchange problem, one has a energy variation between two consecutive measurements. The system starts at a density matrix $\rho_0$, and two consecutive energy measurements are performed, yielding values $E_i$ (at $t=0$), projecting the system at $|i\rangle \langle i|$, and $E_j$ (at $t>0$), both values in the spectrum $H=\{E_1,E_2,...\}$. The energy variation is $\Delta E=E_j-E_i$. In this case, one has a set of energy gaps, $S=\{\Delta E = a-b|a, b \in H\}$, and $P(s)=P(\Delta E)$ given by
\begin{equation}
    P(\Delta E) = \sum_{ij}\delta(\Delta E - (E_j-E_i)) \langle j | \Phi\big(|i\rangle \langle i|\big) |j \rangle \langle i|\rho_0|i\rangle,
\end{equation}
where $\Phi$ is an operator from the quantum dynamical semigroup.
Let the involution $m:S\rightarrow S$ be $m(\Delta E) = -\Delta E$. Check that $m(m(s))=s$. Therefore, $(S,P,m)$ satisfies the condition for the theorem. We have from (\ref{theorem})
\begin{equation}
\label{heattheorem}
    D[P(\Delta E)|P(-\Delta E)] \geq B(w),
\end{equation}
for $B(w)=2w\tanh^{-1}(w)$, with $w$ given by 
\begin{equation}
    w=\frac{1}{2}\sum_{\Delta E} |P(\Delta E)- P(-\Delta E)|.
\end{equation}
Note that, as the theorem for ($S,P,m$) is general, we did not mention thermal distributions or detailed balance in this general setup.

Now for the specific case where the system satisfy the exchange fluctuation theorem (XFT), one has $\Sigma = \alpha \Delta E$ for some affinity $\alpha$ and the XFT \cite{Hasegawa2019,Timpanaro2019B,Evans2002,Merhav2010,Jarzynski2004a} reads
\begin{equation}
\label{xft}
   \frac{P(\Delta E=\Sigma/\alpha)}{P(\Delta E=-\Sigma/\alpha)} = \exp(\Sigma).
\end{equation}
In this case, equation (\ref{heattheorem}) reads
\begin{equation}
\label{heattheorem2}
    \langle \Sigma \rangle \geq B\Big(Prob(\Sigma > 0) - Prob(\Sigma < 0)\Big),
\end{equation}
where we used $\Sigma > 0 \leftrightarrow P(\Delta E = \Sigma/\alpha) > P(-\Delta E = \Sigma/\alpha)$ from (\ref{xft}) and $Prob(\Sigma>0):=\sum P(\Delta E =\Sigma/\alpha)\theta(\alpha \Delta E)$ and $\theta(x)=1 (0)$ if $x>0~(x<0)$ is a step function.

Using that $B'(x)>0$, inverting expression (\ref{heattheorem}) results in a bound for the apparent violations of the second law, $Prob(\Sigma < 0)$, in terms of $\langle \Sigma \rangle$ recently proposed using similar methods \cite{Salazar2021}. The more general bound (\ref{heattheorem}) is still valid without assuming the XFT or detailed balance.

{\bf \emph{Application II: master equation-}}
Another interesting application is to consider continuous Markov process described by a master equation \cite{VanKampen2007}:
\begin{equation}
\label{Markov}
   \dot{p}_i=\sum_{j}W_{ij}p_j-W_{ji}p_i,
\end{equation}

where $W_{ij}\geq 0$ is the transition rate from state $j$ to state $i$ and $p_i=p_i(t)$ is the probability of state $i$, $0\leq p_i (t) \leq 1$, with condition $\sum_{i\neq j} W_{ij}p_j > 0$. Let $S=\{(i,j)|~i,j \in \{1,...,N\}, i\neq j\}$, where we deliberately removed the points $(i,i)$. 

The known expression for the entropy production rate from Schnakenberg \cite{Schnakenberg1976,Tome2012,Ziener2015,Dechant2022} is
\begin{equation}
\label{Schnakenberg}
\overline{\pi}=\frac{1}{2}\sum_{ij}(W_{ji}p_i - W_{ij}p_j)\ln\big(\frac{W_{ji}p_i}{W_{ij}p_j}\big).
\end{equation}
Now we find a natural tuple $(S,P,m)$ for the application of the theorem. For any instant of time,
we define a probability function $P:S\rightarrow [0,1]$ as
\begin{equation}
    P(s)=\frac{W_{ij}p_j}{\sum_{i\neq j}W_{ij}p_j}=\frac{W_{ij}p_j}{z}:=P_{ij},
\end{equation}
for $s=(i,j)$, where $z:=\sum_{i\neq j} W_{ij}p_j$. Note that $P$ is normalized,
\begin{equation}
    \sum_{s\in S}P(s)=
    z^{-1}\sum_{(i,j)\in S} W_{ij}p_j
    =z^{-1}(\sum_{i\neq j}W_{ij}p_j) = 1.    
\end{equation}
Let the involution $m:S\rightarrow S$ be $m(i,j)=(j,i)$. Check that $m(m(s))=s$. Again, $(S,P,m)$ meets the condition for the theorem (\ref{theorem}). Using $P(s)-P(s')=P_{ij}-P_{ji}=(W_{ij}p_j-W_{ji}p_i)/z$, we verify that the total variation $\Delta(P,P')$ is given as
\begin{equation}
    \Delta(P,P')=\sum_{i>j}|J_{ij}|/z = w,
\end{equation}
with $J_{ij}=W_{ij}p_j-W_{ji}p_i$ the probability current. Also, we have
$D(P|P')=\sum_s P(s)\ln(P(s)/P(s'))=\overline{\pi}/z$ as defined in (\ref{Schnakenberg}). In terms of $\overline{\pi}$, $w$ and $z$, theorem (\ref{theorem}) now reads
\begin{equation}
\label{mainresult2}
    \overline{\pi}/z\geq 2w\tanh^{-1}(w),
\end{equation}
which is inequality (\ref{mainresult}), for $\overline{\pi}\rightarrow \overline{\pi}/z$. This bound was recently discovered using other methods \cite{Dechant2022,VanVu2022b}.  In this context, KL and TV are also analyzed in the context of classic speed limit and using Wasserstein distance \cite{Shiraishi2018,VanVu2021}. For consistency of (\ref{mainresult2}), check that $w\leq1$,
\begin{equation}
    w=\frac{1}{z}\sum_{i<j} |W_{ij}p_j-W_{ji}p_i|\leq
    \frac{1}{z}\sum_{i<j}(W_{ij}p_j+W_{ji}p_i) = 1.
\end{equation}

In this case, the bound is 
saturated by a two level system in contact with a thermal bath at infinite temperature. Let $(p_1,p_2)=(p,1-p)$ be the initial state probabilities for some $0.5<p<1$ and transition matrix given by $W_{12}=1$ and $W_{21}=1$, with $W_{11}=W_{22}=0$, 
in the same spirit of (\ref{2lvlb}).

In equilibrium, the system will have $w=0$ (and $\overline{\pi}=0$), satisfying detailed balance condition, $J_{ij}=0$. Near equilibrium, $w\approx 0$, the bound expands to
\begin{equation}
    \label{expansion}
   \overline{\pi}\geq B(w)\approx 2w^2 + \frac{2}{3}w^4 + \mathcal{O}(w^6),
\end{equation}
and recognizing $\overline{\pi} = D(P|P')$, also $\Delta(P,P')=w$, Pinsker's inequality states that $\overline{\pi}=D(P,P')\geq 2\Delta(P,P')^2=2w^2$, so the bound $B(w)$ improves on it in order $\mathcal{O}(w^4)$.

\begin{figure}[htp]
\includegraphics[width=3.3 in]{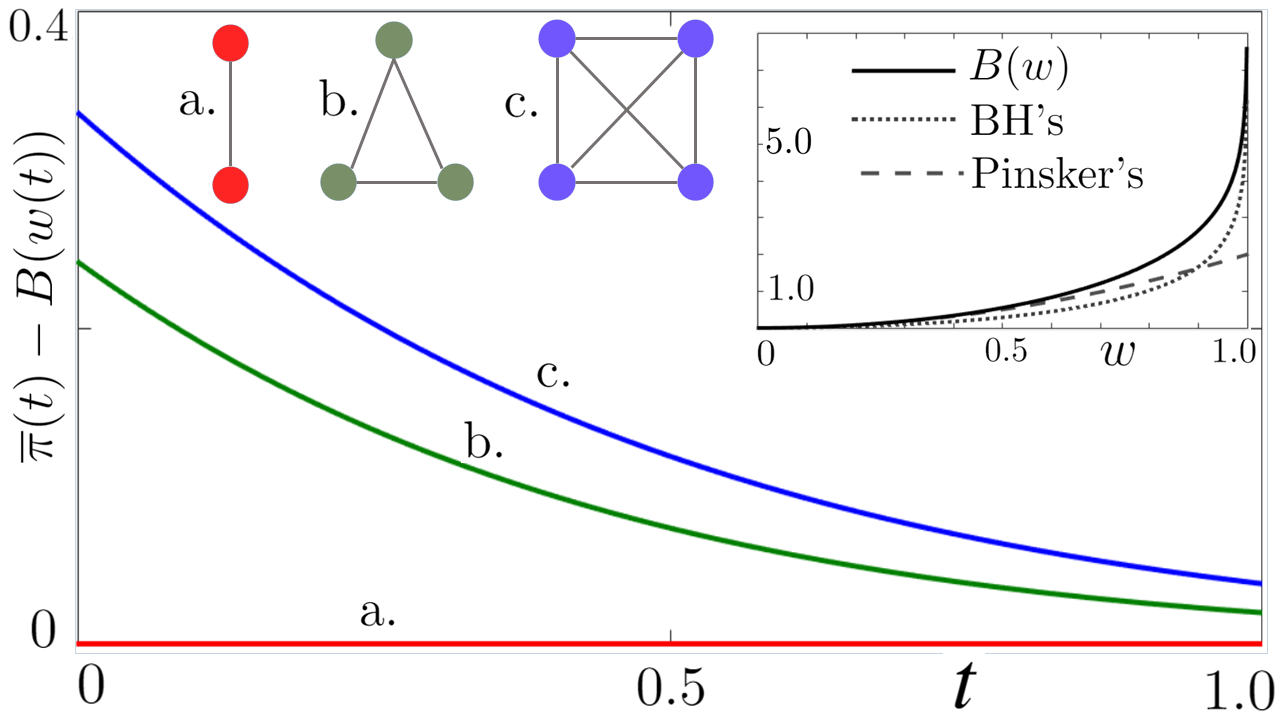}
\caption{(Color online) The difference between the entropy production rate $\overline{\pi}$ and the bound $B(w)$ as a function of time for a two (a.), three (b.) and four-level (c.) systems described in the text. The gaps decay to zero with time as the systems approach equilibrium. The two-level system (a.) saturates the bound. Inset: The bound $B(w)$ as a function of $w$ showing improvement over other known bounds, Pinsker's and Bretagnolle-Huber's (BH's).}
\label{Fig1}
\end{figure}

In Fig.~1, we simulate the master equation (\ref{Markov}) for three different systems: a. a two-level system with $p(0)=[0.9;0.1]$ and $W_{ij}=(1-\delta_{ij})$; b. a three-level system with $p(0)=[0.8;0.1;0.1]$ and $W_{ij}=(1/2)(1-\delta_{ij})$; and c. a four-level system with $p(0)=[0.7;0.1;0.1;0.1]$ and $W_{ij}=(1/3)(1-\delta_{ij})$. They correspond to a weak coupling approximation of a system and a thermal bath with infinite temperature. For each system, we compute $\overline{\pi}(t)$ and $w(t)$ and plot the gap, $\overline{\pi}(t)-B(w(t))$ vs. $t$. Note that the two-level system saturates the bound, as expected. The three and four-level system display a gap that decays in time. In the inset, we observe the plot $B(w)$ vs. $w$, comparing the bound to other known bounds for KL as function of TV. See that Pinsker's is tighter than Bretagnolle-Hubber's (BH's) for $w\approx 0$ (which corresponds to near equilibrium in the Markov dynamics). For $w \approx 1$, the behavior is inverted. In all domain, the figure shows $B(w)$ is a tighter bound than both.

{\bf \emph{Application III: stochastic thermodynamics-}} Consider a trajectory $\Gamma=(x_0,...,x_N)$ and its inverse $\Gamma^\dagger=(x_N,...,x_0)$, where the indexes are time steps and $x_i \in D \subset \mathbb{R}$, representing some domain. The entropy production in stochastic thermodynamics is given as \cite{Seifert2012}
\begin{equation}
\label{prodF}
    \langle \Sigma \rangle_F = \sum_{\Gamma} P_F(\Gamma)\ln\big(\frac{P_F(\Gamma)}{P_B(\Gamma^\dagger)}\big),
\end{equation}
where $P_F$ ($P_B$) is the forward (backward) probability over the set of trajectories. They are different because $P_F$ depends on a protocol $\lambda(t)=(\lambda_0,...,\lambda_N)$, while $P_B$ depends on $\lambda(T-t)=(\lambda_N,...,\lambda_0)$. Define the entropy production for the backward process, $\langle \Sigma \rangle_B$ by replacing $F\leftrightarrow B$ in (\ref{prodF}).

The theorem is applied in this system as follows. Let $S=\{(\Gamma,\sigma)|\Gamma \in D^{N+1}, \sigma \in \{1,-1\}\}$. For any $s=(\Gamma,\sigma)$, define $P(s)=P(\Gamma,\sigma)$ as $P(\Gamma,1):=(1/2)P_F(\Gamma)$ and $P(\Gamma,-1):=(1/2)P_B(\Gamma)$, and check $P(s)$ is normalized in $S$. Let $m(s)=m(\Gamma,\sigma)=(\Gamma^\dagger,-\sigma)$. Check that $m(m(s))=s$.  The tuple $(S,P,m)$ satisfies the condition for the theorem, $D(P|P')\geq B(w)$. In order to understand its meaning, lets compute the KL:
\begin{equation}
    D(P|P')=\sum_{\Gamma,\sigma} P(s)\ln \frac{P(\Gamma,\sigma)}{P(\Gamma^\dagger,-\sigma)}=\frac{\langle \Sigma \rangle_F + \langle \Sigma \rangle_B}{2},
\end{equation}
and the total variation
\begin{equation}
 w=\frac{1}{2}\sum_{\Gamma,\sigma}|P(\Gamma,\sigma)-P(\Gamma^\dagger,-\sigma)|=\frac{1}{2}\sum_\Gamma |P_F(\Gamma)-P_B(\Gamma^\dagger)|,
\end{equation}
so that the theorem reads
\begin{equation}
\label{lowerwork}
    \frac{\langle \Sigma \rangle_F + \langle \Sigma \rangle_B}{2} \geq B(w),
\end{equation}
given in terms of $P_F(\Gamma)$ and $P_B(\Gamma^\dagger)$. For systems where $\langle \Sigma \rangle_{F,B} = \beta (\langle W \rangle_{F,B} - \Delta_{F,B} F):=\beta\langle W_{irr}\rangle_{F,B}$ (irreversible work \cite{Seifert2012}), the theorem results in a lower bound for the sum of forward and backward irreversible works,
\begin{equation}
\label{lowerwork}
    \langle W_{irr} \rangle_F + \langle W_{irr} \rangle_B \geq 2 k_b T B(w),
\end{equation}
for $\beta=(k_b T)^{-1}$ the inverse temperature and $\Delta F$ is the free energy variation.

{\bf \emph{Application IV - time rotation and speed limit-}} Consider a system that outputs stochastic events $s\in S$ for every run with probability $P(s)$ and $m(s)=s'$ is the time reversed event (or any involution), so that $(S,P,m)$ satisfies the theorem. For every run, a demon draws a Bernoulli random variable $\epsilon$ and decides if it plays the dynamics ($\epsilon=1$) or play it backwards ($\epsilon=0$) for that run. Take $\langle \epsilon \rangle = p$ and $0 \leq p \leq 1$. Over several runs, the observer sees the events $s$ with probability $Q(s)$ given by
\begin{equation}
    Q(s)=pP(s)+(1-p)P(s'),
\end{equation}
Note that $(S,Q,m)$ also satisfies the theorem. Over several runs, the observer will measure $D(Q|Q')$, as an attempt to assess the entropy production. In this case, the theorem reads
\begin{equation}
    D(P|P')\geq D(Q|Q') \geq B\big(|2p-1|\Delta(P,P')\big),
\end{equation}
where the first inequality follows from the convexity of KL and the second follows from the theorem, using $Q(s)-Q(s')=(2p-1)(P(s)-P(s'))$. From the theorem, the observer gets a result for $D(Q|Q')$ that is lower bounded by $\Delta(P,P')$ from the original process. This setup works as a time rotation because the demon mixes forward $s$ and backward $s'$ directions. If $p=1/2$, note that $Q(s)=P(s)+P(s')=Q(s')$ and the observer will get $D(Q|Q')=0$, which mimics equilibrium. For some $|2p-1|>0$, $D(Q|Q')$ is lower bounded in terms of the original bound for $D(P,P')$, as $\Delta(P,P')=B^{-1}\circ B(\Delta(P,P'))$ and $B(\Delta(P,P'))$ is the original lower bound (\ref{theorem}).

Alternatively, in another setup, for a set of probabilities $\{P_1,...,P_n\}$ if one has $\sum_i q_i = \tau$, applying the theorem for each $P_i$ results in
\begin{equation}
\sum_i q_i D(P_i|P_i') \geq 2\sum_i q_i w_i \tanh^{-1}(\sum_ j(q_j/\tau)w_i),
\end{equation}
using $B''(x)>0$ again, which can be easily inverted to obtain a lower bound for $\tau$ akin to the classic speed limit, already observed for Markov systems \cite{Dechant2022,VanVu2022b}.

{\bf \emph{Discussion and Conclusions -}} We proved a theorem showing that the KL divergence is bounded by a function of the TV. In this case, we required that the distributions considered are related by an involution, $m(m(s))=s$. This seemly arbitrary condition produced a tighter bound than off the shelf results between KL and TV (Pinsker's and Bretagnolle-Hubber's).

We argue that the involution condition has a physical meaning. It is a general way to introduce the arrow of time: flipping it acts as an involution. Therefore, the events $s$ and $s'$ could be seen as flipped versions of the arrow of time. Because of that, as each event has only one flipped version, the events are able to be collected in pairs $\{s,m(s)\}$ that form the edges of an undirected graph (as well as pairs of the type $\{s,s\}$). This physical constraint allows the discrepancies of $P$ and $P'$, measured in KL and TV, to be tighter than the usual bounds from statistics. Moreover, the theorem translates to a bound between the entropy production rate and the weight of the matching of a graph. The bound is particularly useful, for instance, in situations where the estimation of the weight of the graph is more feasible than the actual entropy production rate.

In this context, the formalism presented in the theorem is general enough to be applied in the derivation of other bounds about the statistics of $P$ and $P'$, when related by an involution, instead of using the fluctuation theorem \cite{Evans2002,Garcia2010,Cleuren2006,Seifert2005,Jarzynski2004a,Andrieux2009,Campisi2015,Merhav2010,Lupos2013,Timpanaro2019B,Hasegawa2019,Neri2017,Pigolotti2017,Yunger2018,Campisi2021,Salazar2021,Deffner2022} as a starting point. For instance, a Thermodynamic Uncertainty Relation (TUR) \cite{Timpanaro2019B,Barato2015,Pietzonka2017,Hasegawa2019} for the variance of $\pi(s)=\ln(P(s)/P'(s))$ or a bound for negative values of $\pi(s)$, $\sum_{\pi(s)<0}P(s)$ \cite{Salazar2021b}, related to apparent violations of the second law as discussed in the applications. In other words, one could formulate some the bounds from nonequilibrium thermodynamics simply in terms of $(S,P,m)$. The extent of this analogy is left for further research.

{\bf \emph{Appendix -}}
For the case $S=\mathbb{R}$, we have $P(s)\rightarrow p(s)ds$ and $\sum_s \rightarrow \int ds$. The distribution $p'$ is given by $p'(s):=p(m(s))|dm(s)/ds|$, the KL divergence is given as $D(p|p')=\int p(s)\ln (p(s)/p'(s))ds$ and TV given as $\Delta(p,p')=(1/2)\int |p(s)-p'(s)|ds$. Now we define another probability density function $f:\mathbb{R}\rightarrow [0,1]$ given by
\begin{equation}
    f(x)=\int_{-\infty}^{\infty}p(s)\delta(\pi(s)-x)ds,
\end{equation}
where $\delta(y)$ is the Dirac delta function and $\pi(s)=\ln (p(s)/p'(s))$ for $p(s)\neq 0$ and $\pi(s)=0$, for $p(s)=0$. We check the normalization,
\begin{equation}
    \int_{x} f(x)dx = \int_{s} p(s) \int_{x}\delta(\pi(s)-x)dxds = \int_{s} p(s)ds = 1,
\end{equation}
and compute the average $\langle x \rangle$,
\begin{eqnarray}
\label{xaverage}
\langle x \rangle := \int_x x f(x) dx 
=\int_{s}\pi(s)p(s)ds= D(p|p').
\end{eqnarray}
The function $f(x)$ satisfies a strong detailed fluctuation theorem (DFT), $f(x)=\exp(x)f(-x)$,
where it follows from
\begin{eqnarray}
f(x)= \int_{s} e^{\pi(s)}p'(s)(\delta(\pi(s)-x)ds\\
=e^x\int_{s'} p(s')\delta(\pi(s')+x)ds'=e^x f(-x),
\end{eqnarray}
We also compute the total variation, $\Delta(f,\hat{f})$, for  $\hat{f}:=f(-x)$, resulting in
\begin{eqnarray}
\label{TV2}
    \Delta(f,\hat{f})=\frac{1}{2}\int_{s}|p(s)-p'(s)|ds=\Delta(p,p').
\end{eqnarray}
Finally, we follow the same steps as in the discrete case (\ref{2lvlb}). Let $\langle x \rangle = \langle |x| \tanh(|x|/2)\rangle$ and $\Delta(f,\hat{f})=\langle \tanh(|x|/2)\rangle$. Using $f_a(x)=(\delta(x-a)\exp(a/2)+\delta(x+a)\exp(-a/2))/(2\cosh(a/2))$, which results in $q(|x|)=\delta(|x|-a)$ as the distribution of $|x|$. In this case, note that
\begin{equation}
\label{saturationinR}
    \langle x \rangle_{f_a} = a\tanh(a/2) = B(\Delta(f_a|\hat{f}_a)),
\end{equation}
for $B(w)=2w\tanh^{-1}(w)$, which means that
\begin{equation}
    D(p|p')=\langle x \rangle = \int_0^\infty  B(\Delta(f_a|\hat{f}_a)) q(a)da \geq B(\Delta(p,p')),
\end{equation}
from Jensen's inequality. Saturation also follows from the two-level system (\ref{saturationinR}).

\bibliography{library}
\end{document}